\documentclass[twoside,fleqn]{article}
\usepackage{times}

\usepackage{lscape} 

\newcommand{\beq}{\begin{equation}}
\newcommand{\eeq}{\end{equation}}

\def\err{\end{array}}
\def\bea{\begin{eqnarray}}
\def\eea{\end{eqnarray}}

\def\inpb{\ifmmode {\rm pb}^{-1}\else ${\rm pb}^{-1}$\fi}
\def\infb{\ifmmode {\rm fb}^{-1}\else ${\rm fb}^{-1}$\fi}
\def\cc{\ifmmode k/\overline M_{Pl}\else $k/\overline M_{Pl}$\fi}
\newskip\zatskip \zatskip=0pt plus0pt minus0pt
\def\matth{\mathsurround=0pt}

\def\atversim#1#2{\lower0.7ex\vbox{\baselineskip\zatskip\lineskip\zatskip
\lineskiplimit 0pt\ialign{$\matth#1\hfil##\hfil$\crcr#2\crcr\sim\crcr}}}
  
\def\pacs#1{\vspace*{-3mm}\begin{quotation}\noindent{\small\sf PACS:} 
{\small #1}
\end{quotation}}
  
\def\title#1{\begin{center}{{\bf #1}}\end{center}\smallskip}
\def\author#1#2{\begin{center}{\bf #1}\\ {\sl #2}\end{center}}
\def\abstract#1{\begin{quotation}\noindent{\small#1}\end{quotation}\medskip}
  
\def\email#1{\footnote{E-mail address: #1}}  

\newenvironment{ack}{%
\vspace{\baselineskip}
\noindent
{\bfseries Acknowledgement:}%
}  

\def\refer#1#2#3#4#5{#1:\ {\sl #2}\ {\bf #3}\ {(#4)}\ #5\ }

\begin{document}

\title{%
ERG AS A TOOL FOR NON-PERTURBATIVE STUDIES\footnote{Talk given 
at the International Conference "Renormalization Group-2002", 
Tatransk\'a \v{S}trba (Slovakia), March 10-16, 2002}}
\author{%
Yuri A. Kubyshin\email{yuri@mat.upc.es}\footnote{On leave of absence 
from the Institute for Nuclear Physics, 
Moscow State University, 119899 Moscow, Russia.}}
{
Departament de Matem\`atica Aplicada IV, 
Universitat Polit\`ecnica de Catalunya \\
C-3, Campus Nord, C. Jordi Girona, 1-3, 08034 Barcelona, Spain}  
\date  
\abstract{Basic elements of the exact renormalization group 
method and recent results within this approach are reviewed. 
Topics covered are the derivation of equations for the effective action 
and relations between them, derivative expansion, solutions of 
fixed point equations and the calculation of the critical exponents, 
construction of the $c$-function and a description of the chiral 
phase transition.}
\pacs{02.50.+s, 05.60.+w, 72.15.-v}
%

\section{Introduction}
\label{sec:intr} 
\setcounter{section}{1}\setcounter{equation}{0} 

To see the role of the exact renormalization group 
approach let us consider a situation which is rather common in 
quantum field theory. 
Suppose that there is a fundamental theory with the action 
$S_{\Lambda_{0}}[\phi]$ defined at some energy scale $\Lambda_{0}$, and  
suppose that we are interested in physical effects described 
by such theory at energies of order $\Lambda \ll \Lambda_{0}$. 
One way to proceed is to calculate perturbatively the renormalization 
group (RG) running of coupling constants and anomalous dimensions from 
$\Lambda_{0}$ down to $\Lambda$. However, 
usually large power and logarithmic corrections develop which make the 
result unreliable. 

Another way is to calculate the low energy 
effective action $S_{\Lambda}[\phi]$ relevant at the 
scale $\Lambda$ and then perform 
the perturbative calculations using this action. The formalism to 
calculate the low energy effective action is the 
Exact Renormalization Group (ERG). It introduces the notion 
of the running action $S_{\Lambda}[\phi]$, i.e. the Wilson effective 
action at scale $\Lambda$, and provides a framework for its 
calculation starting from the initial action 
$S_{\Lambda_{0}}[\phi]$. The basic idea is the following: 
while lowering the scale the effective action $S_{\Lambda}[\phi]$ 
changes in such a way so that to keep the $S$-matrix unchanged. In 
practice a stronger condition is imposed. Namely one requires that 
the Green functions and the generating functional remain unchanged. 
This guarantees that the physical observables do not depend on 
the scale at which the effective action is defined.  

The ERG is a continuous version of the Wilsonian RG \cite{Kad66} adapted 
for applications in quantum field theory. The central element 
of the ERG approach is an equation which determines the change 
of the effective action with the scale. There are quite a few 
ERG equations known and used in practical calculations. 
Below we will consider some of the equations and discuss the relation 
between them. 

The goal of the present contribution is to explain main ideas which 
form the basis of the ERG, outline basic principles of the derivation 
of ERG equations and list a number of characteristic applications of 
this method. We included a discussion of some aspects and features of the 
ERG approach which were not touched or covered sufficiently in 
previous reviews. Nevertheless, many results are left beyond 
the scope of this article. The reader is advised to see, for 
example, Refs. \cite{TM-rev1}-\cite{BTW02} 
for a detailed review and quite complete lists of references to original 
papers. 

The plan of the article is as follows. In Sect. \ref{sec:ERGEs} we 
review the Wegner-Houghton equation as an instructive example, 
describe the general structure of ERG equations and establish the 
relation between them. The Polchinski equation will be discussed 
in some detail. We outline the scheme of calculation of fixed points and 
critical exponents within the ERG approach and present a few examples 
in Sect. \ref{sec:applic}. 
More non-perturbative results will be reviewed in Sect. \ref{sec:applic1}. 
These include the construction of the $c$-function and 
a study of the chiral phase transition. Sect. \ref{sec:concl} contains 
further discussion of general features of the ERG and some concluding 
remarks. 

\section{ERG equations}
\label{sec:ERGEs}

In this article for the sake of simplicity we consider 
almost exclusively the case of the 1 - com\-po\-nent scalar field in 
the $d$-dimensional Euclidean space. 

It is instructive to start with the Wegner-Houghton equation derived in 
Ref. \cite{WH73}. This example illustrates a realization of 
the basic idea, formulated in the Introduction. Let us 
consider the partition function 
\beq
  {\cal Z} = \int \prod_{|p| \leq \Lambda} {\cal D}\phi_{p} 
  e^{-S_{\Lambda}[\phi]},  \label{Z-def}
\eeq
where $\phi_{p}$ is the field in the momentum representation and 
the functional integration is performed over the modes 
with $p$ belonging to the interval $0 \leq |p| \leq \Lambda$. 
The functional $S_{\Lambda}[\phi]$ is the effective action defined 
at scale $\Lambda$. All higher modes are supposed to be integrated out. 
Let $\Lambda'$ be some lower scale, and let us divide the interval 
of momenta into low ($0 \leq p \leq \Lambda'$) and high 
($\Lambda' < p \leq \Lambda$) momentum frequences. Then the 
partition function can be written as 
\beq
  {\cal Z} = \int \prod_{|p| \leq \Lambda'} 
  \prod_{\Lambda' < |p| \leq \Lambda} {\cal D}\phi_{p} 
  e^{-S_{\Lambda}[\phi]} =  
  \int \prod_{|p| \leq \Lambda'} {\cal D}\phi_{p} 
  e^{-S_{\Lambda'}[\phi]},  \label{WH-deriv}
\eeq
so that $S_{\Lambda'}[\phi]$ includes momentum field modes 
with $p$ up to the new scale $\Lambda'$ and is interpreted as 
the effective action at the scale $\Lambda'$. The modes with 
momenta belonging to the shell $\Lambda' < p \leq \Lambda$ have 
been integrated out and led to the corresponding modification of 
the action. This step corresponds to Kadanoff's 
transformation 
(called also blocking or coarsening). As we see, there is a well 
defined boundary between low and high momentum frequences, 
and the latter are integrated out completely. In this case 
it is said that the theory has a sharp momentum cutoff. 

Suppose now that the scales are related by 
$\Lambda' = \Lambda e^{-\delta t}$. 
The authors of Ref. \cite{WH73} showed that if $\delta t \ll 1$, then the  
following relation between the effective actions at these two scales 
fulfills:
\bea
\frac{S_{\Lambda'}[\phi] - S_{\Lambda}[\phi]}{\delta t} 
& = & \frac{1}{2 \delta t} \left\{ \int' dp \left[ 
\ln \frac{\delta^{2} S_{\Lambda}}{\delta \phi_{p} \delta \phi_{-p}} 
- \frac{\delta S_{\Lambda}}{\delta \phi_{p}}
\frac{\delta S_{\Lambda}}{\delta \phi_{-p}} 
\left( \frac{\delta^{2} S_{\Lambda}}
{\delta \phi_{p} \delta \phi_{-p}} \right)^{-1} \right] \right. \nonumber \\
& + & \left.   
(\mbox{rescaling terms}) + {\cal O}\left( \delta t^{2} \right) \right\}, 
\label{WH-eq} 
\eea
where the prime indicates that the integration is performed over 
the shell of momenta $\Lambda' < |p| < \Lambda$.
The origin of the rescaling terms is the canonical change of the 
scale of dimensional parameters due to the change of $\Lambda$. 
Here we prefer to omit these details and focus on the general structure of 
the equation (see Ref. \cite{WH73} and an explicit example below). 
In the limit $\delta t \rightarrow 0$ this relation becomes an 
equation which defines the evolution of 
the effective action with the scale. Being supplied with an initial 
condition 
\[
\left. S_{\Lambda}[\phi] \right|_{\Lambda=\Lambda_{0}} = S_{0}[\phi],
\]
the equation determines the evolution of the effective action with the scale. 
It allows to calculate (in principle) the 
running effective action $S_{\Lambda}[\phi]$ for a given bare action 
("fundamental theory") $S_{0}[\phi]$ defined at $\Lambda_{0} \gg \Lambda$. 

The Polchinski ERG equation is one of the most broadly studied and 
widely used in applications. It was obtained 
in Ref. \cite{Pol84} and is formulated in terms of the Wilson effective 
action with a smooth cutoff function:
\beq
S[\phi;t] = \frac{1}{2} \int \frac{dp}{(2\pi)^{d}} 
\phi_{p} \cdot P^{-1}(p^{2},\Lambda^{2}) \cdot \phi_{-p} + 
S_{int}[\phi;t],  \label{Seff}
\eeq
where
\beq
P(p^{2},\Lambda^{2}) = \frac{1}{p^{2}} K \left( \frac{p^{2}}{\Lambda^{2}} 
\right)    \label{Prop-def} 
\eeq
is a regularized propagator and $K (p^{2}/\Lambda^{2})$ is a (smooth)
ultraviolet cutoff profile (regulating function) which has the following
properties: (1) $K(0) = 1$; (2) $K(z) \rightarrow 0$ as $z \rightarrow
\infty$ fast enough so that all momentum integrals are finite. 
For a smooth cutoff the boundary between low and high momentum frequences 
is blurred, and contributions of high momenta are suppressed rather 
then integrated out. In concrete calculations the exponential 
cutoff function $K(z) = \exp (- a z^{l})$ $(l > 0, \; a > 0)$ is 
often used. The action in the Wegner-Houghton ERG equation 
is regularized with the sharp cutoff function $K(z) = \theta (1-z)$, 
where $\theta$ is the Heaviside (step) function.   
The analysis of the evolution of the effective action with scale 
becomes more transparent if it is carried out in terms of 
dimensionless quantities. For this the dimensionless 
field variable $\varphi_{k}$ and dimensionless momentum $k$ are 
introduced: $\varphi_{k} = \Lambda^{1+d/2} \phi_{p}$, $k = p/\Lambda$. 

In Ref. \cite{Pol84} the Polchinski ERG equation was derived 
starting from the invariance of the partition function under the 
change of the scale (see a detailed discussion of the derivation 
in Ref. \cite{BT94}). In terms of $\varphi_{k}$ and $k$ it 
has the form 
\beq
 \partial_{t} S = {\cal F} \left[ \varphi_{k}, S \right]  
             \label{PERG-gen}  
\eeq
with the functional ${\cal F}$ given by 
\bea
{\cal F}\left[ \varphi_{k}, S \right] & = & 
\int \frac{dk}{(2\pi)^{d}} K'(k^{2}) \left[ 
\frac{\partial S}{\partial \varphi_{k}} 
\frac{\partial S}{\partial \varphi_{-k}} 
  - \frac{\delta^{2} S}{\delta \varphi_{k} \delta \varphi_{-k}}  
  - \frac{2k^{2}}{K(k^{2})} \varphi_{k} 
  \frac{\partial S}{\partial \varphi_{k}}\right] + S d  \nonumber \\
 & + & \int \frac{dk}{(2\pi)^{d}} \left[ 
 \left( 1 - \frac{d}{2} - \frac{\eta}{2} \right) \varphi_{k} 
 \frac{\partial S}{\partial \varphi_{k}} - \varphi_{k} k^{\mu} 
 \frac{\partial'}{\partial k^{\mu}} 
  \frac{\partial S}{\partial \varphi_{k}} \right].   \label{F-PERG}
\eea
The first line contains terms coming from the "blocking transformation", 
the terms in the second line are "rescaling terms".  

One can check that 
changing the cutoff $K(k^{2})$ can be compensated by a field 
redefinition \cite{LatMor}. This explains why approximations of equations 
for different cutoff functions give very close results. Moreover, certain 
equations can be transformed one into another by field redefinitions 
\cite{LatMor} (see also \cite{Weg74}). They form a "universality" class of 
equivalent ERG equations and determine the same low energy effective actions. 
In particular, one can show that the Polchinski ERG equation is equivalent 
to the equation for the Legendre effective action 
\cite{Mo-LERG94} or to the equation for the average effective action 
derived in Ref. \cite{We93} (see also \cite{Paw02}). 

The ERG equations were derived and extensively studied for scalar 
(see Sect. \ref{sec:applic1}) and spinor theories \cite{CKM95}. 
A non-invariant formalism for gauge theories was developed and applied 
in a number of papers (see for example Refs. \cite{Paw02}, \cite{gauge}, 
\cite{gauge1}). There the gauge invariance, not preserved by the 
equation, was restored at the end by imposing certain conditions on the 
RG flow. A gauge invariant formalism has been recently proposed in 
Refs. \cite{Mo98} (see Refs. \cite{AKMT}, \cite{Ar02} for recent 
developments). 

\section{Fixed points and critical exponents}
\label{sec:applic}

Let us first discuss the type of problems 
which can be naturally addressed within the ERG approach and then 
review some results. 

An ERG equation can be written in general form (\ref{PERG-gen}) 
with some functional ${\cal F}$ (see, for example, Eq. (\ref{F-PERG})). 
Search for fixed point (FP) solutions $S^{*}[\phi]$ is one of the immediate 
applications of the ERG. A FP is described by the action  
satisfying the condition $\partial_{t}S^{*}=0$ or, equivalently, the 
equation 
\beq
  {\cal F} [\varphi, S^{*}] = 0.   \label{FP-eq}
\eeq
FP solutions do not contain any scale and represent continuum 
limits of the theory. 

Suppose that we have found a FP solution $S^{*}[\phi]$. The next problem 
to address is to study a linearized theory around the FP \cite{Weg72}. 
For this one writes the effective action as 
$S[\varphi;t] = S^{*}[\varphi] + \Delta S$ and expands 
Eq. (\ref{PERG-gen}): 
\beq
  \partial_{t} \Delta S  = {\cal F}[\varphi, S^{*} + \Delta S] = 
  {\cal L} \cdot \Delta S + {\cal O} (\Delta S^{2}),  \label{lin-exp}
\eeq
where ${\cal L} \cdot \Delta S$ is the part linear in $\Delta S$. The 
operator ${\cal L}$ defines a set of eigenoperators $\Phi_{n}[\varphi]$ 
and critical (or scaling) exponents $\lambda_{n}$ \cite{GolRied76}: 
\[
{\cal L} \Phi_{n}[\varphi] = \lambda_{n} \Phi_{n}[\varphi].
\]
As a result the effective action in the vicinity of the FP $S^{*}$ 
is equal to 
\[
S[\varphi;t] = S^{*}[\varphi] + \sum_{n} \mu_{n} e^{\lambda_{n}t} 
\Phi_{n}[\varphi].
\]
If $\lambda_{n} > 0$, $\lambda_{n} < 0$ or $\lambda_{n} = 0$ 
the associated operator $\Phi_{n}[\varphi]$ and 
its coupling $\mu_{n}e^{\lambda_{n}t}$ are called relevant, irrelevant 
or marginal, respectively. Relevant operators have couplings  
which grow along the flow, and they alone are responsible 
for scaling effects near the FP. The 
couplings of the irrelevant operators decrease along the flow,  
such operators contribute to subleading corrections. To determine the 
behavior of a marginal operator higher order terms in the expansion 
of the action around the FP must be taken into account. The critical 
exponents are physical observables and can be measured experimentally. 

The ERG equations are quite complicated equations in functional 
derivatives with respect to the field variable $\varphi$ and in ordinary 
derivatives with respect to the flow parameter $t$. The formal exactness 
of the ERG by itself does not constitute any calculational progress compared 
to the perturbation theory. It is the existence of non-perturbative 
approximation schemes which opens possibilities for exploration of 
non-perturbative physics and makes, therefore, the approach 
quite valuable. 

The most widely used scheme is the derivative expansion \cite{Myer75}. 
The main idea is to expand the action of interaction in powers 
of space-time derivatives of the field: 
\beq
S_{int} = \int d^{d}x \left[ V(\phi (x);t) + 
\frac{1}{2} (\partial_{\mu} \phi)^{2} Z(\phi (x);t) + \ldots \right],  
                  \label{deriv-exp}
\eeq
where potentials $V(\phi;t)$, $Z(\phi;t)$, etc. do not contain derivatives.
Note that here the action is written in terms of the fields in the 
coordinate representation.  
The leading-order (LO) approximation is obtained by retaining only 
the first term in (\ref{deriv-exp}) and neglecting all derivatives in the 
effective action \cite{NiChSt74} - \cite{HH86}. 
It is called the local potential approximation (LPA). 
Substituting expansion (\ref{deriv-exp}) into 
the Polchinski ERG equation, Eqs. (\ref{PERG-gen}), (\ref{F-PERG}), 
we obtain a system of coupled partial differential equations 
for $V(\phi;t)$, $Z(\phi;t)$, etc. 

As an example let us consider the next-to-leading order (NLO)  
approximation. Denote the field variable by $z$ and introduce the function 
$f(z;t) \equiv \partial V(z;t)/\partial z$. The system of equations 
becomes \cite{BHLM94}  
\bea
\partial_{t} f & = & - A f'' -  2BZ' + 2K'(0) f f' + 
\left( 1 + \frac{d}{2} - \frac{\eta}{2} \right) f + 
\left( 1 - \frac{d}{2} - \frac{\eta}{2} \right) z f,  \label{NLO-1}  \\
\partial_{t} Z & = & - A Z'' + 2K''(0) (f')^{2} + 4K'(0) Z f' + 
2 K'(0) f Z' \nonumber \\
& + & \left( 1 - \frac{d}{2} - \frac{\eta}{2} \right) z Z' - 
\eta Z - \frac{\eta}{2}. \label{NLO-2}
\eea
Here the prime denotes the derivative with respect to the field variable 
$z=\varphi$ and $A$, $B$ stand for the following integrals: 
\[
A = \int \frac{dk}{(2\pi)^{d}} K'(k^{2}), \; \; \; 
B = \int \frac{dk}{(2\pi)^{d}} k^{2} K'(k^{2}). 
\]

The case of 1-component scalar field was studied in detail in a 
number of papers, see for example Refs. \cite{HH86} - \cite{Mo-ERG}. 
The FP solutions are given by system (\ref{NLO-1}), (\ref{NLO-2}) 
with $f$ and $Z$ independent of the flow parameter $t$, i.e. 
$\partial_{t} f = 0$, $\partial_{t} Z = 0$ in the l.h.s.    
Let us consider the theory with $Z_{2}$-symmetry under 
the reflection $\phi \rightarrow -\phi$. The boundary conditions 
fixing a solution are usually chosen at $z=0$ in the form 
\[
 f(0) = 0,  \; \; \; f'(0) = \gamma, \; \; \; 
    Z(0) = 0,  \; \; \; Z'(0) = 0. 
\]
The first and the last conditions follow directly from the 
$Z_{2}$-symmetry (recall that $f(z)$ is the derivative of the potential).  
The third relation is a normalization condition,   
meaning that the second term in (\ref{deriv-exp}) does not 
contribute to the massless kinetic term in (\ref{Seff}).

In the LPA the Polchinski ERG equation reduces to the following equation 
for $f(z)$:  
\beq
A f'' =  2K'(0) f f' + \left( 1 + \frac{d}{2} - 
\frac{\eta}{2} \right) f + 
\left( 1 - \frac{d}{2} - \frac{\eta}{2} \right) z f.  \label{NLO-FP1}
\eeq
which follows from Eq. (\ref{NLO-1}). The consistency of approximation 
requires $\eta=0$. 
Eq. (\ref{NLO-FP1}) is stiff, and for a general value of $\gamma$ the 
solution is singular at some finite value $z_{0}(\gamma)$ of the 
field variable. Of course, such solution does not give sensible 
potential and, therefore, is not physical. By fine tuning 
the parameter $\gamma$ to a value $\gamma = \gamma_{*}$ such that 
$z_{0}(\gamma_{*}) = \infty$ one obtains a physical FP solution. By 
spanning all values of $\gamma$ one obtains all FP solutions accessible 
in the LPA  \cite{FelFil}, \cite{Mo94a}. 

A more general study which gives a deeper insight into the 
structure of the space of solutions of Eq. (\ref{NLO-FP1}) was 
carried out in \cite{KNR01}, \cite{KNR02} (also for the case of 
$N$-component scalar theory). It was shown that if arbitrary 
$\eta$ are allowed, then regular solutions form a discrete set of 
families corresponding to curves $\eta_{n}(\gamma)$ in the 
$(\gamma,\eta)$-plane, where $n=1,2, \ldots$ labels the curves. 
They are described by the formula 
\beq
   \eta_{n} (\gamma) = 2 + d \alpha_{n} \left( \frac{\gamma}{d} \right), 
     \label{eta-n}
\eeq
where $\alpha_{n}(z)$ are universal functions which do not depend on 
the space-time dimension and can be calculated from Eq. (\ref{NLO-FP1}). 
Their properties were studied in Ref. \cite{KNR02}. In particular, 
$\alpha_{n} (0) = - n/(n+1)$ and $\alpha_{n} \rightarrow -1$ 
for $z \rightarrow -\infty$ so that the curves accumulate at $\eta=2-d$. 

The curves of regular solutions for various $d$ follow the universal 
pattern given by Eq. (\ref{eta-n}). When we pass from one number 
of dimensions to another the pattern shifts vertically and the 
curves scale in the $\gamma$-direction. The physical FPs correspond to 
the values of $\gamma=\gamma_{*}$ at which $\eta_{n}(\gamma)$ cross 
the $\gamma$-axis. This explains correctly the number of FPs in 
various dimensions $d > 2$. For $d=2$ the curves approach 
asymptotically the line $\eta=0$, this is an indication of the 
existence of infinite number of FPs in this case. 

FP solutions were studied using the ERG first in Ref. \cite{HH86} for 
the sharp cutoff Wegner-Houghton equation, and later for the other 
ERG equations in the LO and NLO approximations, see for example 
Refs. \cite{BHLM94}, \cite{Mo95} - \cite{KNR98}. In all these 
and other similar studies consistent results were obtained: 

1) In $d=4$ dimensions the only FP found is the trivial Gaussian FP. 

2) In $d=3$ there are just two FPs: 
the Gaussian FP and a non-trivial FP. 
Calculation of the critical exponents confirm that the latter 
corresponds to the universality class of the Wilson-Fischer FP 
(Ising model universality class). 

3) In $d=2$ the FP corresponding to the $p(p+1)$ conformal models 
were found \cite{Mo95}, \cite{KNR98}. 

The critical exponents to the LO and NLO of the derivative expansion were 
calculated in a number of papers. In Table 1  
we present a few examples 
of numerical results for the Wilson-Fischer FP in $d=3$. These include 
the anomalous dimension $\eta$, the correlation length critical 
exponent $\nu = 1/\lambda_{1}$, associated to the relevant operator, 
and $w = - \lambda_{2}$ for the least irrelevant operator. Note that 
$\eta = 0$ in the LPA. In the case when the exponent under consideration 
depends on the choice of the cutoff function $K$ in Eq. (\ref{Prop-def}) 
the intervals of values corresponding to certain ranges of the regulator 
parameter are indicated. This is analogous to the scheme dependence 
in the perturbative RG. One can see that the LO and NLO of the 
derivative expansion give fairly good results.  
It turns out that the values obtained within the Polchinski 
ERG equation (see the Table) coincide with those for the optimal 
regulator within the average effective approach \cite{Lit02}. 

\begin{table}[t]
\label{tab1}
\begin{tabular}{|l|c|c|c|}
\hline 
Approach and approximation & $\eta$ & $\nu$ & $w$ \\
\hline 
Wegner-Houghton eq., ${\cal O}(\partial^{0})$ \cite{HH86} 
& 0 & 0.687 & 0.595 \\
\hline
Polchinski eq. \cite{BHLM94}& & & \\
${\cal O}(\partial^{0})$ & 0 & 0.649 & 0.66 \\
${\cal O}(\partial^{2})$ & 0.019-0.056 & 0.616-0.637 & 0.70-0.85 \\
\hline
Legendre effective action eq. \cite{Mo-ERG}, \cite{Mo95} - \cite{Mo97} & & & \\
${\cal O}(\partial^{0})$ & 0 & 0.660 & 0.628 \\
${\cal O}(\partial^{2})$ & 0.054 & 0.618 & 0.897 \\
\hline
World best estimates & 0.035(3) & 0.631(2) & 0.80(4) \\
\hline
\end{tabular}
\caption{Results of calculations of the anomalous dimension $\eta$ and 
critical exponents $\nu$ and $w$ for the Wilson-Fischer FP in $d=3$.
${\cal O}(\partial^{0})$ and ${\cal O}(\partial^{2})$ denote the 
LPA and NLO approximation of the derivative expansion respectively.  
The entries of the last row (taken from the article by Morris \cite{Mo96a}) 
were obtained by averaging the world best estimates \cite{Z-J74}.}
\end{table}

The ERG equations were also used for the calculation of the RG flows 
of the effective action \cite{Mo97}, \cite{HKLM94}, \cite{Wet91}, 
\cite{BTW96} and the Ising model equation of state 
\cite{Mo97}, \cite{BTW96}, \cite{BerWet97}, as well as for the analysis  
of phase transitions \cite{SeWe99} (see also \cite{BTW02}).   
These studies show high efficiency and accuracy of the ERG approach. 
Thus, as it was observed in Ref. \cite{TM-rev2}, the comparison 
with the results obtained within the perturbation 
theory and within the $\epsilon$-expansion shows that 
while the perturbative methods are more accurate for low 
order couplings, the derivative expansion eventually overtakes 
for higher orders. 

\section{More non-perturbative results}
\label{sec:applic1}

\subsection {$c$-function} 

Another class of non-perturbative results obtained within the 
ERG is the construction of the $c$-function. 
The $c$-function appears in the context of the $c$-theorem first proved 
by Za\-mo\-lod\-chi\-kov~\cite{Zam86} for 2-dimensional theories. 
A proof of the theorem in four 
dimensional AdS space was given in Ref. \cite{FoLa98}. 

The $c$-theorem states 
that for Poincar\'e invariant, renormalizable and unitary theories 
there exists a non-negative 
function of the flow parameter $t$, which was called the $c$-function, such  
that: (1) it decreases along RG trajectories, i.e. $dc / dt < 0$; 
(2) it is stationary at a FP: $dc /dt|_{F.P.} = 0$. 

Considered with respect to a local (in the space of all possible 
interactions) basis of operators and corresponding couplings $\{g^{i}\}$, 
the $c$-function depends on $t$ only through $g^{i}(t)$: $c = c(g^{i}(t))$. 
The total derivative is equal to 
$dc / dt = - \beta^{i} \partial c / \partial g^{i}$, where 
$\beta^{i}$ are $\beta$-functions. The $c$-theorem 
expresses the irreversibility of RG flows, which is a manifestation of 
the decoupling of massive states as the system flows towards the infrared 
limit and provides a valuable instrument to relate realizations of the 
same quantum system at different scales. 

Moreover, the RG flow is gradient, i.e. for a 
chosen basis of operators 
\[
\beta^{i} \equiv -\frac{dg^{i}}{dt} = \sum_{j} {\cal G}^{ij} 
\left( \{ g \} \right)
\frac{\partial c}{\partial g^{j}}, 
\] 
where ${\cal G}^{ij}$ is a positive definite metric (known as the 
Zamolodchikov metric) in the space 
of couplings. This property of the $c$-function means that only FPs are 
allowed in the space of couplings, and limit cycles or more complicated 
behaviors are excluded. 

One of the first studies of the $c$-function within the ERG approach was 
carried out in Ref. \cite{HKLM94} within the Wegner-Houghton formulation. 
The $c$-function within the Polchinski ERG formalism in the 
LO of the derivative expansion was studied in Ref. \cite{GHFMo97}. 
It is based on the observation that LPA equation (\ref{NLO-FP1}) 
(with $\eta=0$) can be cast into the form \cite{Zum94} 
\[
 a G(\phi) \frac{\partial \rho}{\partial t} = - 
 \frac{\delta {\cal D}[\rho]}{\delta \rho}, 
\]
where
\bea
& & G(\phi) = e^{- (d-2)\phi^{2}/4}, \; \; \; 
\rho (\phi;t) = e^{-V(\phi;t)}, \nonumber \\
& & {\cal D}[\rho] = a \int d\phi G(\varphi) \left[ \frac{1}{2} 
\left( \frac{d \rho}{d \phi} \right)^{2} + 
\frac{d}{4} \rho^{2} (1 - 2 \ln \rho)  \right], \nonumber 
\eea
and $a$ is a normalization factor. 
It was found that in terms of $G$ and ${\cal D}$ the 
$c$-function and the metric can be written as follows \cite{GHFMo97}: 
\[
c(\{g\}) = \frac{1}{A} \ln \left( \frac{4 {\cal D}}{d} \right), \; \; \;  
{\cal G}_{ij} = a {\cal D} \ln A \int d\phi G \rho^{2} 
{\cal O}_{i}(\phi) {\cal O}_{j}(\phi),  
\]
where $O_{i}(\phi) = \partial V/\partial g^{i}$ and $A$ is another 
normalization factor. 

\subsection{Quark-meson transition}

Systems with many degrees of freedom are often described by different 
relevant excitations (fields) at different 
scales. Thus, in the theory of strong interactions at 
$\Lambda^{-1} \ll 1\;$fm the relevant degrees of freedom are quarks and 
gluons and the adequate theory is the QCD. At $\Lambda^{-1} \gg 1\;$fm 
the observed particles are mesons and hadrons, and their 
interaction is described by the chiral model (nonlinear $\sigma$-model). 

A formalism to tackle the change of relevant degrees of freedom 
within the ERG approach was developed and used for the analysis of 
phenomena like the chiral phase transition in Refs. \cite{EllWet94}, 
\cite{JW96} (see also \cite{BTW02}). Here we just outline the main 
elements of this formalism. The idea is to use the ERG flow equation 
for a model of quarks and gluons for scales $\Lambda > \Lambda_{\sigma}$, 
and a correspondingly modified ERG equation for quarks and mesons 
for $\Lambda < \Lambda_{\sigma}$, where $\Lambda_{\sigma}$ is the transition 
scale ($\sim 600-700$MeV). 
At $\Lambda = 1.5\;$GeV the system was described by a QCD inspired 
model which symbolically can be written as  
\[
\Gamma_{\Lambda} [\psi,\bar{\psi}] = \int \frac{dq}{(2\pi)^{4}} 
 \bar{\psi}_{q} \hat{q} \psi_{-q} + \frac{1}{2} \int \prod_{i=4}^{4} 
 \frac{dq_{i}}{(2\pi)^{4}} \gamma_{\Lambda} (q_{i})   
 (2\pi)^{4} \delta (\sum q_{i}) 
 \mu (q_{i}) (\bar{\psi} \psi) \; (\bar{\psi} \psi),  
\]
where $\psi$, $\bar{\psi}$ denote the quark fields and 
$(\bar{\psi} \psi) \; (\bar{\psi} \psi)$ stands for different 
4-quark interactions allowed by vector and axial vector symmetries. 
The gluons are supposed to be integrated out, and the form of the 
4-quark couplings is motivated by the gluon exchange and 
a confining potential. It turns out that, as the integration of 
modes is performed and the scale lowers down, the effective action 
develops a pole-like structure of a bound state: 
\[
\gamma_{\Lambda} \sim -g(q_{1},q_{2}) \tilde{G}(s) g(q_{3},q_{4}),
\] 
where $g(q_{1},q_{2})$ corresponds to the Bethe-Salpeter wave function and 
$\tilde{G}(s)$ is the bound-state propagator with a pole-like dependence 
on $s$. At the scale $\Lambda = \Lambda_{\sigma} = 0.63\;$GeV 
another effective action was introduced which symbolically 
could be written as 
\beq
\Gamma_{\Lambda} [\psi,\bar{\psi},\sigma] =\Gamma_{\Lambda} [\psi,\bar{\psi}]
+ \frac{1}{2} O^{+} \tilde{G} O - \sigma^{+} O + 
\frac{1}{2}\sigma^{+} \tilde{G}^{-1} \sigma,     \label{QCD-action2}
\eeq
where $O$, $O^{+}$ are composite operators quadratic in $\psi$, $\bar{\psi}$. 
They are appropriately defined so that the 
pole-like structure cancels. $\sigma$, $\sigma^{+}$ are 
collective fields playing the role 
of mesons of the linear $\sigma$-model with Yukawa couplings. 
Action (\ref{QCD-action2}) is used as the initial condition 
for the running effective action at lower scales. Further integration 
of the ERG equation gives the meson potential which develops a non-trivial 
minimum and thus describes the spontaneous breaking of chiral symmetry at 
$0 \leq \Lambda < \Lambda_{\phi}$. From this the authors of Ref. 
\cite{EllWet94} calculated the vacuum expectation value  
$<\sigma> = 0.18\;$MeV and the chiral condensate
$<\psi \bar{\psi}> \approx 
(175 \; \mbox{MeV})^{3}$. 
A substantial discrepancy between the calculated value of the pion decay 
constant and its experimental value was explained by the crudeness 
of the approximation \cite{EllWet94}. A more systematic study 
of the quark-meson system within the average effective action 
approach with gluon effects taken into account is presented in 
\cite{JW96}.

\section{General Remarks and Conclusions}
\label{sec:concl}

{}From the numerous studies it can be concluded that the ERG 
is a powerful method in quantum field theory. 
Its effectiveness and reliability in practical 
calculations has been confirmed in numerous 
applications in various models. In cases when 
the results can be compared with other methods or with experimental data 
it was observed that the ERG gives a fairly good precision 
already at the LO or NLO of the derivative expansion. 
We would like to mention that even at the level of the LPA 
the effective potential $V(\phi;t)$ is given by quite a non-trivial 
expression which does not rely on expansion in powers of the 
field or any small parameter and includes Feynman diagrams of 
all orders and topologies \cite{Mo-LERG94}. 

In the present article we do not discuss the 
relation between the ERG and the perturbative RG. 
We restrict ourselves to a few comments. In principle, 
perturbative calculations can be carried out 
within the ERG approach and reproduce known results. 
For this one performs the perturbative expansion in the Polchinski ERG 
equation, Eqs. (\ref{PERG-gen}), (\ref{F-PERG}), and 
resolves the resulting 
system of equations, for example, by iterations. This amounts to calculation 
with a smooth momentum cutoff regulator $K(p^{2}/\Lambda^{2})$ in  
Eq. (\ref{Prop-def}). Needless to say that this regularization technique 
for performing perturbative computations is is not among the efficient ones. 
A detailed discussion of the relation between the ERG approximations, 
e.g. the derivative expansion, and the perturbation theory expansion, 
issues of the scheme dependence, renormalization conditions, etc. 
can be found, for example, in Refs. \cite{Kub98}, \cite{MoT99}, \cite{LiPa02}, 
\cite{Gat02}. 

The advantage of the ERG is that it allows approximation schemes 
which are not based on a small parameter expansion. Perhaps the 
most successful of them is the derivative expansion discussed in detail 
in Sect. \ref{sec:applic}. A weak point of this technique is the absence 
of an obvious parameter or a practical criteria controlling 
its convergence. A complete and consistent study of 
its convergence and accuracy is still missing. 
In a number of papers the polynomial approximation of the effective 
potential in powers of the field was used \cite{HKLM94}, \cite{MOP}. 
Though in certain cases such truncations give rather good numerical 
results, the procedure is not convergent and generates spurious solutions 
\cite{Mo94a}. 
Development of other non-perturbative approximations is highly 
necessary for addressing a wider class of physical problems, especially 
those which require different degrees of freedom 
at different scales. We feel that the ERG approach, being 
exact by construction, allows for such approximations. 

Let us mention another important feature of the ERG. 
In the space of theories one may introduce locally, i.e. in a vicinity 
of a given FP, a basis of operators ${\cal O}_{i}$ and a set of 
corresponding coupling constants $\{ g^{i}(\Lambda\}$ which 
play the role of local coordinates (masses are also treated as couplings). 
Some of the couplings will be relevant or marginal (we denote them 
by $\{ \tilde{g}^{i} (\Lambda) \}$), others will be irrelevant (see 
Sect. \ref{sec:applic}). Using the ERG equation one can show that 
the running effective action can be expressed as a self-similar flow 
of a finite number of relevant couplings: 
\beq
S_{\Lambda}[\phi] = S[\phi; \tilde{g}(\Lambda)] \left( 1 + {\cal O} \left( 
\frac{\Lambda}{\Lambda_{0}} \right) \right). 
\label{self-sim}
\eeq 
The irrelevant couplings bring only corrections of order 
$\Lambda/\Lambda_{0}$ \cite{TM-rev1}, \cite{BT94}. 
The limit $\Lambda_{0} \rightarrow \infty$ 
is often called the continuum limit in the literature. Property 
(\ref{self-sim}) suggests that there should be certain similarity 
between the ERG and the RG approach based on the exact 
functional equation by Bogoliubov and Shirkov \cite{BSh55} 
(see \cite{Sh96} for a review). Namely, in both cases the 
underlying symmetry is the functional self-similarity, 
i.e. the invariance of the solution (running effective action) under 
the choice of the initial condition (bare action $S_{0}$). 
Studying this relation in more detail and sharing techniques 
of the functional RG equation approach \cite{KovSh01} and the ERG could 
be mutually beneficial.
If one considers a system with an external classical field $H$ then all 
reasonable interactions of the quantum field $\phi$ with $H$ can be 
incorporated into certain couplings of the effective action, 
i.e. we pass from a self-similar flow 
$S[\varphi; \tilde{g}(\Lambda)]$ to $S[\phi; \tilde{g}(\Lambda, H)]$. 
This observation is in full accordance with the result obtained 
in Ref. \cite{Kaz02} in the framework of the perturbative RG.

\begin{ack} We are grateful to  D.I. Kazakov, J.I. Latorre, D. Litim, 
T.R. Morris, J. Pawlowski and D.V. Shirkov for numerous discussions 
and useful remarks. We wish to thank the organizers of the Conference 
"Renormalization Group-2002" for providing an extremely stimulating 
environment. The financial support from the Polytechnical University 
of Catalonia and from the grants CERN/P/FIS/40108/2000 
and "Universities of Russia" is acknowledged.    
\end{ack}

\end{document}